\documentclass[aps,pra,superscriptaddress,10pt,tightenlines,nofootinbib]{revtex4}
\usepackage{amsmath,amsthm,graphicx,amssymb,verbatim}
\usepackage[usenames, dvipsnames]{color}
\usepackage[ pdftex, plainpages = false, pdfpagelabels,
                 pdfpagelayout = useoutlines,
                 bookmarks,
                 bookmarksopen = true,
                 bookmarksnumbered = true,
                 breaklinks = true,
                 linktocpage=all,
                 pagebackref=false,
                 colorlinks = true,
                 linkcolor = BrickRed,
                 urlcolor  = blue,
                 citecolor = BrickRed,
                 anchorcolor = green,
                 hyperindex = true,
                 hyperfigures
                 ]{hyperref}
\usepackage{xifthen} 

\newcommand{\tr}{\hbox{tr}}

\newcommand{\ket}[1]{{\ensuremath{\left| #1 \right\rangle}}}
\newcommand{\bra}[1]{{\ensuremath{\left\langle #1 \right|}}}
\newcommand{\braket}[2]{{\ensuremath{\left\langle #1 \middle| #2
      \right\rangle}}}
\newcommand{\ketbra}[2]{{\ensuremath{\left| #1 \middle\rangle \middle\langle #2
      \right|}}}
\newcommand{\hilb}{\mathcal{H}}
\newcommand{\cF}{\mathcal{F}}
\newcommand{\fourier}[1]{\cF\left(#1\right)}
\newcommand{\inprod}[2]{\left\langle #1 , #2 \right\rangle}

\newcommand{\arxiv}[2][]{\ifthenelse{\isempty{#1}}{\href{http://arxiv.org/abs/#2}{{\tt arXiv:\allowbreak{}#2}}} {\href{http://arxiv.org/abs/#2}{{\tt arXiv:\allowbreak{}#2 [#1]}}}}

\newcommand{\bq}{\begin{quotation}}
\newcommand{\eq}{\end{quotation}}

\newcommand{\ECgp}{\hbox{EC}}

\begin{document}

\title{The SIC Question: History and State of Play}
\author{Christopher A.\ Fuchs}
\affiliation{Physics Department, University of Massachusetts Boston}
\author{Michael C.\ Hoang}
\affiliation{Computer Science Department, University of Massachusetts Boston}
\author{Blake C.\ Stacey}
\affiliation{Physics Department, University of Massachusetts Boston}
\date{\today}

\begin{abstract}
Recent years have seen significant advances in the study of symmetric
informationally complete (SIC) quantum measurements, also known as
maximal sets of complex equiangular lines.  Previously, the published
record contained solutions up to dimension 67, and was with high
confidence complete up through dimension 50.  Computer calculations
have now furnished solutions in all dimensions up to 151, and in
several cases beyond that, as large as dimension 844.  These new
solutions exhibit an additional type of symmetry beyond the basic
definition of a SIC, and so verify a conjecture of Zauner in many new
cases. The solutions in dimensions 68 through 121 were obtained by
Andrew Scott, and his catalogue of distinct solutions is, with high
confidence, complete up to dimension 90. Additional results in
dimensions 122 through 151 were calculated by the authors using
Scott's code.  We recap the history of the problem, outline how the
numerical searches were done, and pose some conjectures on how the
search technique could be improved.  In order to facilitate
communication across disciplinary boundaries, we also present a
comprehensive bibliography of SIC research.
\end{abstract}

\maketitle

\section{Introduction}

The problem of symmetric, informationally complete quantum
measurements~\cite{Zauner:1999, Caves:2002, Renes:2004, Scott:2010a}
stands at the confluence of multiple areas of physics and mathematics.
SICs, as they are known for short, tie into algebraic number
theory~\cite{Appleby:2016, Bengtsson:2016, FOOP:2017, Appleby:2017},
higher-dimensional sphere packing~\cite{Stacey:2016}, Lie and Jordan
algebras~\cite{Appleby:2011a, Appleby:2015}, finite
groups~\cite{Zhu:2015a, Stacey:2016b} and quantum information
theory~\cite{Fuchs:2010, Tabia:2012, Tabia:2013, Fuchs:2013, Zhu:2012,
  Graydon:2016, Zhu:2016a, Szymusiak:2015, Appleby:2016b,
  Fuchs:2016a}.  Without the study of SICs, one might think that the
intersection of all these subjects would have to be the empty set.
And yet, for all that, a SIC is a remarkably simple mathematical
structure, as structures go.  Consider the complex vector space
$\mathbb{C}^d$.  To a physicist, this is the Hilbert space associated
with a $d$-level quantum system.  Let $\{\ket{\psi_j}\}$ be a set of
exactly $d^2$ unit vectors in~$\mathbb{C}^d$ such that
\begin{equation}
\left|\braket{\psi_j}{\psi_k}\right|^2
 = \frac{1}{d+1}
\label{eq:SIC}
\end{equation}
whenever $j \neq k$.  The set $\{\ket{\psi_j}\}$, which can be
associated with a set of pairwise equiangular lines through the
origin, is a SIC.

One can prove that no more than $d^2$ vectors in a $d$-dimensional
Hilbert space can be equiangular.  That is, if $\{\ket{\psi_j}\}$ is a
set of vectors, and $\left|\braket{\psi_j}{\psi_k}\right|^2 = \alpha$
for every $j \neq k$, then that set can have at most $d^2$ elements.
In addition, for a maximal set the value of~$\alpha$ is fixed by the
dimension; it must be $1/(d+1)$.  So, a SIC is a maximal equiangular
set in~$\mathbb{C}^d$; the question is whether they can be constructed
for \emph{all} values of the dimension.  Despite a substantial number
of exact solutions, as well as a longer list of high-precision
numerical solutions~\cite{Scott:2010a, Scott:2017, Appleby:2017}, the
problem remains open.

Exact solutions, found by hand in a few cases and by computer algebra
software in the others, are known in the following dimensions:
\begin{equation}
d = 2\hbox{--}24, 28, 30, 31, 35, 37, 39, 43, 48, 124.
\end{equation}
The historical record of exact solutions has been spread over several
publications~\cite{Scott:2010a, Chien:2015b, Appleby:2017,
  Grassl:2017}.  For several years, the most extensive published set
of numerical results went as high as dimension $d =
67$~\cite{Scott:2010a}.  Now, numerical solutions are known in all
dimensions up to and including $d = 151$, as well as a handful of
other dimensions up to~$d = 844$.  These numerical solutions were
found using code designed and written by Scott, who extended the
results of~\cite{Scott:2010a} through $d = 121$ using his personal
computer over several years of dedicated effort.  In addition, Scott
found solutions in a set of dimensions ($d = 124$, 143, 147, 168, 172,
195, 199, 228, 259, 323) by taking advantage of particular simplifying
assumptions that are applicable in those dimensions~\cite{Scott:2017}.
Further close study of these properties led to a solution for $d =
844$~\cite{Grassl:2017}.  Because dimension $d = 121$ was pushing the
limits of what was computationally feasible without those simplifying
assumptions, the authors calculated solutions in dimensions 122
through 151 by running Scott's code on the Chimera supercomputer at
UMass Boston.  In turn, Scott was able to employ another algorithm
(outlined below) to refine the numerical precision of these results.

The solutions from all of these search efforts are available together
at the following website:
\begin{quotation}
\url{http://www.physics.umb.edu/Research/QBism/}
\end{quotation}

An intriguing feature of the SIC problem is that some numerical solutions,
if extracted to sufficiently high precision, can be converted to exact
ones~\cite{Chien:2015b, Appleby:2017}.  Most recently, this technique
was used to derive an exact solution in dimension $d = 48$.  Another
interesting aspect is that the number of distinct SIC constructions
varies from one dimension to another.  (The sense in which two SICs
can be equivalent will be discussed in detail below.)  One reason
computational research is valuable, beyond extending the list of
dimensions in which SICs are known, is that it provides what is likely
a complete picture for many values of the dimension.  This is
important for understanding the subtle connection between SICs and
algebraic number theory~\cite{Appleby:2016}, a connection that brings
a new angle of illumination to Hilbert's twelfth
problem~\cite{Bengtsson:2016, FOOP:2017}.

SICs are so called because, thanks to the rules of quantum theory, a
SIC in~$\mathbb{C}^d$ specifies a measurement procedure that can, in
principle, be applied to a $d$-level quantum system.  For example, a
SIC in~$\mathbb{C}^2$ is a set of four equiangular lines, and it is a
mathematical model of a measurement that a physicist can perform on a
single qubit.  The term ``informationally complete''---the ``IC'' in
``SIC''---means that if one has a probability distribution for the
possible outcomes of a SIC experiment, one can compute the
probabilities for the possible outcomes of any other experiment
carried out on the target system~\cite{Fuchs:2013}.  So, while one can
pose the question of their existence using pure geometry, SICs are
relevant to applied physics.  Indeed, SIC measurements have recently
been performed or approximated in the laboratory~\cite{Du:2006,
  Durt:2008, Medendorp:2011, Pimenta:2013, Bian:2014, Zhao:2015,
  Bent:2015, SosaMartinez:2017}, and they are known to be optimal
measurements for quantum-state tomography~\cite{Scott:2006}.

A SIC provides a \emph{frame}---more specifically, an
\emph{equiangular tight frame}---for the vector space $\mathbb{C}^d$.
Given a finite-dimensional Hilbert space $\hilb$ with an inner product
$\inprod{\cdot}{\cdot}$, a frame for~$\hilb$ is a set of vectors
$\{v_j\} \subset \hilb$ such that for any vector $u \in \hilb$,
\begin{equation}
A ||u||^2 \leq \sum_j \left|\inprod{v_j}{u}\right|^2
 \leq B ||u||^2,
\end{equation}
for some positive constants $A$ and $B$.  The frame is
\emph{equal-norm} if all the vectors $\{v_j\}$ have the same norm, and
the frame is \emph{tight} if the ``frame bounds'' $A$ and $B$ are
equal.  The ratio of the number of vectors to the dimension of the
space is known as the \emph{redundancy} of the
frame~\cite{Et-Taoui:2016}.  For more on this terminology and its
history, we refer to Kova\v{c}evi\'c and Chebira~\cite{Kovacevic:2007,
  Kovacevic:2007b}.  In our experience, the language of frames is more
common among those who come to SICs from pure mathematics or from
signal processing than among those motivated by quantum physics.

Any vector in~$\mathbb{C}^d$ can be represented by its inner products
with all the SIC vectors.  In quantum physics, one also considers the
set of Hermitian operators on~$\mathbb{C}^d$.  This set in fact forms
a Hilbert space itself, with a dimension of~$d^2$, and the inner
product given by the Hilbert--Schmidt formula
\begin{equation}
\inprod{A}{B} = \tr(AB).
\end{equation}
Rewriting the SIC vectors $\{\ket{\psi_j}\}$ as rank-1 projection
operators,
\begin{equation}
\Pi_j = \ketbra{\psi_j}{\psi_j},
\end{equation}
we construct a nonorthogonal basis for the Hilbert space of Hermitian
operators.  Because the inner products of these projectors are
uniform, given by
\begin{equation}
\tr(\Pi_j\Pi_k) = \frac{d\delta_{jk} + 1}{d+1},
\end{equation}
then it is straightforward to find a shifting and rescaling that
orthogonalizes the basis $\{\Pi_j\}$, at the cost of making the
operators non-positive-semidefinite.  In fact, there are two
choices:
\begin{equation}
Q_j^\pm = \pm \sqrt{d+1}\,\Pi_j
 + \frac{1 \mp \sqrt{d+1}}{d} I.
\end{equation}
The bases $\{Q_j^\pm\}$ have interesting properties with regard to Lie
algebra theory~\cite{Appleby:2015} and the study of quantum
probability~\cite{Zhu:2016a, Fuchs:2017}.

\section{Generating SICs with Groups}

All known SICs have an additional kind of symmetry, above and beyond
their definition: They are \emph{group covariant.}  Each SIC can be
constructed by starting with a single vector, known as a
\emph{fiducial} vector, and acting upon it with the elements of some
group.  It is not known in general whether a SIC must be group
covariant.  Because such an assumption greatly reduces the search
space~\cite{Scott:2010a, Appleby:2016}, it has been the only method
used so far: The fact that we only know of group-covariant SICs
could potentially be an artifact of this.  (However, we do have a
proof that all SICs in~$d = 2$ and $d = 3$ are group
covariant~\cite{Hughston:2016}.)

In all cases but one, the group that generates a SIC from a fiducial
is an instance of a \emph{Weyl--Heisenberg group.}  We can define this
group as follows. First, fix a value of~$d$, and let $\omega = e^{2\pi
  i / d}$. Let $\{\ket{0}, \ket{1}, \ldots, \ket{d-1}\}$ be an
orthonormal basis for the Hilbert space $\hilb_d =
\mathbb{C}^d$. Then, construct the shift and phase operators
\begin{equation}
X\ket{j} = \ket{j+1},\ Z\ket{j} = \omega^j \ket{j},
\end{equation}
where the shift is modulo $d$.  These operators satisfy the
Weyl commutation relation,
\begin{equation}
X^l Z^\alpha = \omega^{-l\alpha}Z^\alpha X^l.
\end{equation}
In a sense, the operators $X$ and $Z$ come as close as possible to
commuting, without actually doing so:  The only cost to exchanging
their order is a phase factor determined by the dimension.

The Weyl--Heisenberg \emph{displacement operators} in dimension $d$
are defined by
\begin{equation}
D_{l\alpha} = (-e^{i\pi / d})^{l\alpha} X^l Z^\alpha.
\end{equation}
The product of two displacement operators is, up to a phase factor, a third:
\begin{equation}
D_{l\alpha}D_{m\beta} = (-e^{i\pi/d})^{\alpha m - \beta l} D_{l+m,\alpha+\beta}.
\end{equation}
Therefore, by allowing the generators to be multiplied by phase
factors, we can define a group, known as the \emph{Weyl--Heisenberg
  group} in dimension $d$.  This group dates back to the early days of
quantum physics.  Weyl introduced the generators $X$ and $Z$ as long
ago as 1925 in order to define what one might mean by the quantum
theory of discrete degrees of freedom~\cite{Weyl:1931, Scholz:2006,
  Scholz:2007} (see also \cite[pp.\ 2055--56]{Fuchs:2014}).  This
group, and structures derived from it, are critically important in
quantum information and computation; for example, this is the basic
prerequisite for the Gottesman--Knill theorem, which indicates when a
quantum computation can be efficiently simulated
classically~\cite{Gottesman:1999}.  The close relationship between
SICs and the Weyl--Heisenberg group suggests that SICs are a kind of
structure that quantum physics should have been studying all along.

Zhu has proved that in prime dimensions, group covariance implies
Weyl--Heisenberg covariance~\cite{Zhu:2010}.  The one known exception
to the rule of Weyl--Heisenberg covariance is the \emph{Hoggar
  SIC}~\cite{Hoggar:1981, Hoggar:1998}, which lives in a prime-power
dimension, $d = 8$.  As in all other dimensions, there is a
Weyl--Heisenberg SIC, but there is \emph{also} the Hoggar SIC\@.  Like
many other exceptions to mathematical classifications, it is related
to the octonions~\cite{Stacey:2016, Stacey:2016b}.

One example of a Weyl--Heisenberg SIC can be constructed by taking the
orbit of the following two-dimensional vector under the
Weyl--Heisenberg displacement operators:
\begin{equation}
\ket{\psi_{0}^{\rm(qubit)}} = \frac{1}{\sqrt{6}} \left(\begin{array}{c}
 \sqrt{3 + \sqrt{3}} \\
 e^{i\pi/4} \sqrt{3 - \sqrt{3}}
 \end{array}\right).
\end{equation}
This orbit is a set of four vectors.  In the Bloch sphere
representation, they form the vertices of a regular tetrahedron
inscribed within the sphere.

An example in dimension $d = 3$, one which is remarkable for the
further subtle symmetries it possesses beyond even group covariance, is the
orbit of
\begin{equation}
\ket{\psi_{0}^{(\rm Hesse)}} = \frac{1}{\sqrt{2}}
\left(\begin{array}{c} 0 \\ 1 \\ -1
 \end{array}\right)
\label{eq:hesse-fiducial}
\end{equation}
under the Weyl--Heisenberg displacements.  This set of vectors is
known as the \emph{Hesse SIC}~\cite{Hughston:2007, Dang:2013,
  Hughston:2016}, thanks to its relation with the Hesse configuration
familiar from design theory and the study of cubic curves,
specifically nonsingular cubic curves in complex projective
two-space~\cite{Coxeter:1991, Tabia:2013, Zhu:2015a, Stacey:2016c}.

\section{Historical Overview}

In order to understand the current state of SIC research, one must
grasp how people came to the SIC question, what other structures they
think are related, what tools they suspect are applicable, and so
forth.  A physicist, motivated by quantum information theory, is apt
to have a different mental context than a pure mathematician driven by
the abstract appeal of geometry.  To attempt to foster an
interdisciplinary discussion, we provide in this section a brief
historical overview.

The Hesse SIC can be extracted from a 1940 article by
H.\ S.\ M.\ Coxeter~\cite{Coxeter:1940}, which discusses what he later
called the ``Hessian polyhedron''~\cite{Coxeter:1991}.  This
polyhedron lives in three-dimensional complex vector space and has 27
vertices (which correspond to the 27 lines on a cubic
surface~\cite{Baez:2016}).  The vectors in the Hesse SIC are a subset
of those vertices.  Mathematicians were studying SICs in $d = 2$ and
$d = 3$ more explicitly as early as the 1970s, in the terminology of
complex equiangular lines~\cite{Delsarte:1975}.  Stuart Hoggar found a
SIC in dimension $d = 8$ a few years later~\cite{Hoggar:1981}, by
considering the diagonals of a quaternionic polytope and converting
their coordinates to complex numbers.  The SICs in $d = 2$ and $d =
3$, together with the Hoggar lines in $d = 8$, still stand out among
the known SICs; various unusual attributes they possess have led them
to be designated the \emph{sporadic SICs}~\cite{Stacey:2016,
  FOOP:2017}.

In a 1987 article, Richard Feynman used a construction that is in
retrospect a $d = 2$ SIC to study the probability theory of a qubit.
SICs entered quantum theory more generally starting with the work of
Gerhard Zauner, who began to consider the problem in the
1990s~\cite[p.\ 1941]{Fuchs:2014}.  By 1999, Zauner had found the
connection with the Weyl--Heisenberg group and proven SIC existence up
to $d = 5$~\cite{Zauner:1999}.  He also posed a conjecture that the
search for Weyl--Heisenberg SICs could be simplified by considering a
particular unitary operator~\cite{Zauner:1999}, a conjecture we will
describe in detail below.

Independently of Zauner, Carlton Caves developed the idea of a SIC as
representing a quantum measurement~\cite{Caves:2002}, motivated
originally by attempts to prove the ``quantum de Finetti
theorem''~\cite{Caves:2002c}.  SICs turned out to have more symmetry
than was necessary for that proof, but they soon took on a life of
their own in quantum information theory~\cite{Fuchs:2016a}.  The term
``SIC'' (first pronounced as ``sick,'' and later like ``seek'') dates
to this period.  A 2004 paper, ``Symmetric Informationally Complete
Quantum Measurements''~\cite{Renes:2004}, introduced them to the
mainstream of the quantum information community, and reported
numerical solutions for dimensions up to $d = 45$.  In a later survey,
Andrew Scott and Markus Grassl extended these numerical results up
to dimension $d = 67$~\cite{Scott:2010a}. Further work by Scott
established Weyl--Heisenberg SICs in all cases up to $d = 121$ without
exception.  Moreover, these results together are probably a complete
catalogue of all distinct Weyl--Heisenberg SICs up to dimension $d =
90$~\cite{Scott:2017}.  In later sections, we will give an overview of
how these searches were done.

Exact solutions were found more slowly.  Having exact expressions for
SIC fiducial vectors allowed Appleby~\emph{et al.}\ to discover a
connection with Galois theory~\cite{Appleby:2013}.  In turn, this led
to further relations with algebraic number theory, a frankly mysterious
development that is still under active
investigation~\cite{Appleby:2016, FOOP:2017, Bengtsson:2016, Appleby:2017}.

An exhaustive treatment of SIC research would require transgressing
many disciplinary boundaries, and would doubtless grow to an
intimidating length.  In order to provide at least a helpful ordering
on the literature, we have made our bibliography as comprehensive as
possible.  We note in particular a selection of papers that address
SICs in the context of abstract algebra~\cite{Appleby:2011a, Zhu:2014,
  Appleby:2015, Appleby:2017}, algebraic number
theory~\cite{Appleby:2013, Appleby:2016, FOOP:2017, Bengtsson:2016},
category theory~\cite{Wetering:2017}, finite group
theory~\cite{Zhu:2010, Appleby:2012b, Bengtsson:2012b, Zhu:2015a,
  Szymusiak:2015, Stacey:2016, Stacey:2016b, Chien:2017}, quantum
computing and contextuality~\cite{Bengtsson:2012, Cabello:2012,
  Zhu:2016a, Stacey:2016b, Fuchs:2016b, Howard:2016, Fuchs:2017,
  Andersson:2017}, and quantum entanglement~\cite{Chen:2015,
  Shen:2015, Graydon:2016, Graydon:2016a}.  We also note the
significant number of student theses written in whole or in part on
the SIC problem~\cite{Zauner:1999, Renes:2004a, MKF:2008,
  YadsanAppleby:2012, Zhu:2012, Tabia:2013b, Andersson:2014,
  Blanchfield:2014, Chien:2015b, Dang:2015, Stacey:2015}.

Before moving on, we note that the \emph{real} analogue of the SIC
problem, i.e., finding maximal sets of equiangular lines in real
vector spaces, has also been of considerable interest to
mathematicians~\cite{Sloane:2015, Greaves:2016, Balla:2016}.  The
maximal number of equiangular lines in a $d$-dimensional vector space
is not $d^2$, but only $d(d+1)/2$.  That is, if we have a set of $N$
unit vectors $\{\hat{v}_i\}$ in a $d$-dimensional vector space, such
that
\begin{equation}
|\langle \hat{v}_i, \hat{v}_j \rangle| = \alpha\ \forall\ i \neq j,
\label{eq:real-inprod}
\end{equation}
then the size $N$ of the set cannot exceed $d(d+1)/2$.  Moreover,
while the complex bound of $d^2$ has been saturated in every dimension
that we have been able to check, it is known that the real bound of
$d(d+1)/2$ is not even attained for all values of~$d$.  For example,
in $d = 7$, one can construct a set of $7 \cdot 8 / 2 = 28$
equiangular lines, but this is also the best that can be done in $d =
8$.  In fact, the only known instances where the bound of $d(d+1)/2$
can be attained are dimensions 2, 3, 7 and 23~\cite{Balla:2016}.

There is a sign freedom in this definition of the angle, since
Eq.~(\ref{eq:real-inprod}) is satisfied if the inner product $\langle
\hat{v}_i, \hat{v}_j\rangle$ is either $+\alpha$ or $-\alpha$.  The
presence of this discrete choice means that investigations of real
equiangular lines often have a rather combinatorial flavor.  In
contrast, when we take the magnitude of a complex inner product, we
discard a continuous quantity, a phase that in principle can be
anywhere from~$0$ to $2\pi$.  Generally speaking, the ``feel'' of the
real and complex problems differ, as is evidenced by the different
areas of mathematical expertise brought to bear upon them.  However,
subtle and unanticipated points of contact between the real and
complex cases do exist~\cite{Stacey:2016b}.

\section{How to Search for SICs Numerically}

As before, let $\{\ket{j}\}$ be an orthonormal basis for the Hilbert
space $\hilb_d = \mathbb{C}^d$.  In this basis, the fiducial vector
can be written
\begin{equation}
\ket{\psi_0} = \sum_j a_j \ket{j},
\end{equation}
for some set of coefficients $\{a_j\}$.

Acting with the Weyl--Heisenberg operator $D_{l\beta}$ on the fiducial
vector $\ket{\psi_0}$ produces a new vector, whose squared innner
product with the fiducial vector is
\begin{equation}
[{\bf F}]_{\beta l} = \left|\bra{\psi_0} X^l Z^\beta \ket{\psi_0}\right|^2
 = \left|\sum_j a^*_j a_{j+l} \omega^{-j\beta}\right|^2.
\end{equation}
The right-hand side has the form of the magnitude squared of a Fourier
coefficient, \emph{i.e.,} of a power spectrum.  Specifically, the set
of squared inner products between~$\ket{\psi_{(l,\beta)}}$ and
$\ket{\psi_0}$ for any given value of~$l$ is the power spectrum of the
sequence
\begin{equation}
f^{(l)}_j = a^*_j a_{j+l}.
\end{equation}
By the Wiener--Khinchin theorem, we know that the power spectrum of a
sequence is the Fourier transform of the autocorrelation of that
sequence~\cite{Wiener:1964}.  Therefore, the autocorrelation of the
sequence $f^{(l)}_j$, when put through the Fourier transform, will
yield the sequence $[{\bf F}]_{\beta l}$.  The set of autocorrelation
sequences for all values of~$l$ forms a matrix.  Using $\star$ to
denote the correlation of two sequences, we can write the elements of
this matrix as
\begin{equation}
[{\bf G}]_{kl} = (f^{(l)}\star f^{(l)})_k
 = \sum_j a_j a^*_{j+k} a^*_{j+l} a_{j+k+l}.
\label{eq:G-from-f}
\end{equation}
The matrix ${\bf G}$ is in many situations more convenient to work
with than the original matrix ${\bf F}$, because ${\bf G}$ lacks phase
factors and treats both of its indices on equal footing.  For example,
it is apparent from the definition of~${\bf G}$ that
\begin{equation}
[{\bf G}]_{kl} = [{\bf G}]_{lk}.
\end{equation}
This is equivalent to a property of the matrix ${\bf F}$ that is less obvious
to the eye:
\begin{equation}
[{\bf F}]_{lm} = \frac{1}{d} \sum_{\alpha,\beta}
  \omega^{-\alpha l + \beta m} [{\bf F}]_{\beta \alpha}.
\label{eq:F-self-inverse}
\end{equation}

If we take the Fourier transform of the columns of the matrix ${\bf
  G}$,
\begin{equation}
\fourier{\{[{\bf G}]_{kl}\}}_\beta
 = \sum_k \omega^{-k\beta} [{\bf G}]_{kl} ,
\end{equation}
we recover the squared inner products between the candidate SIC
vectors and the fiducial.  This means that if the vectors
$\{\ket{\psi_{(l,\beta)}}\}$ really do comprise a SIC, then the
matrix~${\bf G}$ must take a very specific form.  Every entry
in~$\fourier{\{[{\bf G}]_{kl}\}}_\beta$ must equal $1/(d+1)$, except for the
element at~$l = \beta = 0$, which equals 1.  Recalling that a constant
sequence is the discrete Fourier transform of a Kronecker delta
function, we can deduce the desired values of~$[{\bf G}]_{kl}$.

The result is that if $\ket{\psi}$ is a Weyl--Heisenberg fiducial
vector, then
\begin{equation}
[{\bf G}]_{kl} = \sum_j a_j a^*_{j+k} a^*_{j+l} a_{j+k+l}
 = \frac{\delta_{k0} + \delta_{l0}}{d+1}.
\label{eq:fiduciality}
\end{equation}
This implication also works in reverse, thanks to the transitivity of
the group action.

The basic idea of finding SICs numerically is to use standard
optimization methods to find a fiducial vector that makes $[{\bf G}]_{kl}$ as
close to the desired form as possible.  Note that, when ${\bf G}$ is
constructed from a SIC fiducial,
\begin{equation}
\sum_{k,l} |[{\bf G}]_{kl}|^2 = \frac{2}{d+1}.
\end{equation}
One can prove~\cite{Appleby:2014b} that this is a lower bound.  In
general,
\begin{equation}
\sum_{k,l} |[{\bf G}]_{kl}|^2 \geq \frac{2}{d+1},
\label{eq:fiduciality2}
\end{equation}
and the inequality is saturated if and only if the input vector is truly a
SIC fiducial.

This naturally suggests a way to find SIC fiducial vectors: Minimize
the LHS of the inequality in Eq.~(\ref{eq:fiduciality2}), aiming for
the lower bound given on the RHS.  During our time investigating the
SIC question, we have at various points implemented this idea in
Mathematica, in Python and in C++ using the GNU Scientific Library.
We find in general that numerical optimization finds a local minimum
quickly, but a local minimum might only imply inner products between
the vectors that are correct to a few decimal digits.  A way around
this problem is to repeat the optimization many times, starting from
different points in the search space.  Since these trials can run
concurrently, the problem is amenable to parallelization.  This is the
approach we followed when using the Chimera supercomputer to obtain
solutions in dimensions 122 through 151.  Scott's implementation,
which we employed on Chimera, uses a C++ code for a limited-memory
quasi-Newton optimization algorithm, L-BFGS, due to
Liu~\cite{Liu:2012}.

\begin{figure}[h]
\includegraphics[width=12cm]{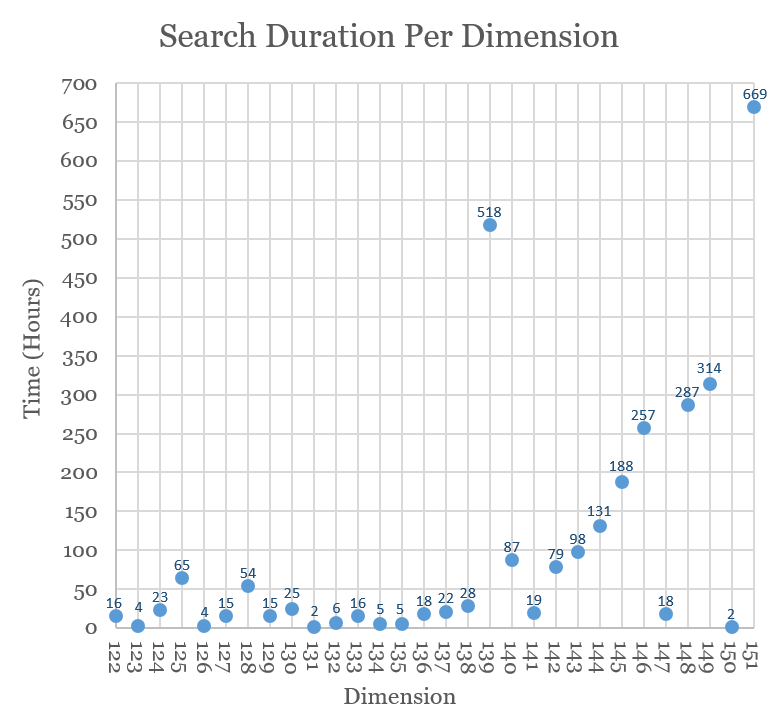}
\caption{\label{fig:searchtimes} Time (in hours) spent searching for a
  Weyl--Heisenberg SIC in dimensions 122 through 151.}
\end{figure}

As is evident from Figure~\ref{fig:searchtimes}, the time required to
obtain solutions did not increase steadily with the dimension.  For
example, $d = 146$ took eleven days of computer time and $d = 148$
required twelve days, but $d = 147$ took only 18 hours.  Likewise,
Chimera spent 28 days trying to find a $d = 151$ solution before
succeeding, but it found a SIC in $d = 150$ in only two hours.  (These
figures are all for ``wall clock'' elapsed time.  The number of
processor-hours devoted was greater, since we ran Scott's code in
parallel on 96 of Chimera's cores.)  We suspect the variation is due
to different numbers of inequivalent solutions existing in different
dimensions: The more solutions, the easier it is to hit upon one of
them.

Once we have a numerical result in hand, we can refine its precision.
This requires a code that uses multi-precision arithmetic, which will
run more slowly than the optimization in the first
step~\cite{Scott:2017}.  The fiducial vectors available
at~\cite{Scott:2017} and at the website referenced above were obtained
in this way and are accurate to 150 digits.

Before moving on, we note a conjecture, based on numerical evidence,
that hints at additional hidden structure in the SIC problem.  Note
that the definition of ${\bf G}$ implies
\begin{equation}
\left|[{\bf G}]_{k,l}\right|
 = \left|[{\bf G}]_{-k,l}\right|
 = \left|[{\bf G}]_{k,-l}\right|
 = \left|[{\bf G}]_{-k,-l}\right|
 = \left|[{\bf G}]_{l,k}\right|
 = \left|[{\bf G}]_{-l,k}\right|
 = \left|[{\bf G}]_{l,-k}\right|
 = \left|[{\bf G}]_{-l,-k}\right|.
\label{eq:redundancies}
\end{equation}
Here, indices are to be interpreted modulo $d$.  Because any
autocorrelation attains its maximum at zero offset, we also know
immediately that the elements of~${\bf G}$ cannot be larger in the
middle of the matrix than they are on the edges:
\begin{equation}
\left|[{\bf G}]_{kl}\right| \leq [{\bf G}]_{0l}.
\end{equation}
The Fourier transform of $[{\bf G}]_{kl}$ over the index $k$ is, by the
Wiener--Khinchin theorem, the power spectrum of~$f^{(l)}$.  Because
power spectra are nonnegative, we can say that
\begin{equation}
[{\bf G}]_{-k,l} = [{\bf G}]_{k,l}^*.
\label{eq:G-cc}
\end{equation}

Are there additional symmetries or redundancies, not so apparent from
the definition?  By happenstance, one of the authors (CAF) observed
that imposing a \emph{subset} of the constraints in
Eq.~(\ref{eq:fiduciality}) was sufficient to find a SIC fiducial
vector~\cite[pp.\ 1252--59]{Fuchs:2014}.  Specifically, by finding a
solution to the simultaneous equations
\begin{align}
[{\bf G}]_{0k} &= \frac{\delta_{k0} + 1}{d+1}, \\
[{\bf G}]_{1k} &= \frac{\delta_{k0}}{d+1}, \\
[{\bf G}]_{2k} &= \frac{\delta_{k0}}{d+1},
\end{align}
one finds a solution to \emph{all} the equations in
(\ref{eq:fiduciality}).  The redundancies in
Eq.~(\ref{eq:redundancies}) are sufficient to imply that this holds up
to $d = 5$.  We call the idea that it remains true in all dimensions
the ``$3d$ conjecture.''  It has been verified numerically up to
dimension $d = 28$~\cite{Appleby:2014b}.  If the $3d$ conjecture is
indeed true, it would reduce the complexity of the problem, as
measured by the number of simultaneous equations to solve, from
\emph{quadratic} in the dimension to \emph{linear.}

\section{Zauner Symmetry}

Is there any way to narrow the search space for SIC fiducial vectors?
To see how to answer this in the affirmative, we must elaborate upon
the group theory we discussed in the previous sections.  The
\emph{Clifford group} for dimension $d$ is the ``normalizer'' of the
Weyl--Heisenberg group: It is the set of all unitary operators that,
acting by conjugation, map the set of Weyl--Heisenberg operators in
dimension $d$ to itself.  We saw earlier how the orbit of a vector
under the Weyl--Heisenberg group can be a SIC; likewise, we can study
the orbit of a vector under the entire Clifford group.

For our purposes in this note, the important point is that if we
conjugate a Weyl--Heisenberg operator $D_{kl}$ by a Clifford unitary
$U$, we obtain a Weyl--Heisenberg operator $D_{k'l'}$, possibly with an
additional phase.  Details on the construction and representation of the
Clifford group in any finite dimension $d$ can be found in
Appleby~\cite{Appleby:2005b}.

It was conjectured by Zauner~\cite{Zauner:1999}, and independently by
Appleby, that in every dimension $d > 2$, a Weyl--Heisenberg SIC
fiducial exists that is an eigenvector of a certain order-3 Clifford
unitary, which is now known as the \emph{Zauner unitary.}  Acting on
the Weyl--Heisenberg generators, the Zauner unitary effects the change
\begin{equation}
X^mZ^n \mapsto X^{-n} Z^{m-n},
\label{eq:zauner-once}
\end{equation}
up to an overall phase factor.  (See equation (3.10b) of
Zauner~\cite{Zauner:1999}, or equation (127) of
Appleby~\cite{Appleby:2005b}.)  Applying this again yields
\begin{equation}
X^{-n} Z^{m-n} \mapsto X^{n-m} Z^{-n-m+n} = X^{n-m}Z^{-m},
\label{eq:zauner-repeat}
\end{equation}
and a third iteration gives
\begin{equation}
X^{n-m} Z^{-m} \mapsto X^{m}Z^{n-m+m} = X^m Z^n,
\end{equation}
confirming that this operation has order 3.

How might assuming the Zauner conjecture simplify the search for SICs?
First, we will make some remarks on this from an algebraic
perspective, and then we will address the point in a way suited to
numerical optimization.  Let $\ket{\psi}$ be a candidate fiducial
vector, and suppose that it is an eigenvector of the Zauner unitary
$U$ with unit eigenvalue:
\begin{equation}
U\ket{\psi} = \ket{\psi}.
\end{equation}
Consequently,
\begin{equation}
\left|\bra{\psi} X^nZ^m \ket{\psi}\right|^2
 = \left|\bra{\psi}U^\dag X^n Z^m U \ket{\psi}\right|^2.
\end{equation}
As $U$ is a Clifford unitary, requiring that $\ket{\psi}$ is an
eigenvector of~$U$ implies degeneracies among the elements of the
matrix ${\bf F}$.

Because
\begin{equation}
X^a \mapsto Z^a \mapsto X^{-a} Z^{-a},
\end{equation}
the Zauner unitary sends the left edge of ${\bf F}$ to the top edge
and then to the main diagonal.  More generally, specifying a column
of~${\bf F}$ (which is equivalent to fixing a column of~${\bf G}$) and
imposing the Zauner condition means that the same constraint also
simultaneously fixes a row and a diagonal.

Earlier, in Eq.~(\ref{eq:F-self-inverse}), we saw that a symmetry
of~${\bf G}$ implied a Fourier-type relation among the elements
of~${\bf F}$.  We have expressed the Zauner condition as a degeneracy
within~${\bf F}$, but what does it imply for~${\bf G}$?  The result
can be found straightforwardly, and it closely resembles
Eq.~(\ref{eq:F-self-inverse}):
\begin{equation}
[{\bf G}]_{kl} = \frac{1}{d} \sum_{\alpha,\beta}
  \omega^{k\alpha + l\beta} [{\bf G}]_{\beta,\alpha - l}.
\label{eq:G-self-inverse}
\end{equation}
This the expression of Zauner symmetry in the ${\bf G}$ matrix.  A
special case of note: If we set $k = l = 0$, then
\begin{equation}
[{\bf G}]_{00} = \frac{1}{d}\sum_{\alpha,\beta} [{\bf G}]_{\beta,\alpha}
 =  \frac{1}{d}\sum_{\alpha,\beta} [{\bf G}]_{\alpha,\beta}.
\end{equation}

We note that the assumption that the fiducial is a Zauner eigenvector
is enough to prove some additional cases of the $3d$ conjecture, up to
dimension $d = 9$, in a straightforward way.  In dimension $d = 5$,
the basic symmetries of ${\bf G}$ imply that the $3d$ constraints
automatically specify the entire matrix ${\bf G}$, and thus also fix
${\bf F}$ to have the desired form for a SIC.  By imposing the
condition that our initial vector is a Zauner eigenvector, we can
extend this up to dimension $d = 8$.  This can be seen directly by
drawing an $8 \times 8$ grid and shading in the appropriate squares.

In fact, we can carry this argument a little further.  We obtain ${\bf
  G}$ by Fourier transforming the columns of~${\bf F}$.  Therefore, if
we specify $\{[{\bf G}]_{k0}\}$, we automatically fix $\{[{\bf F}]_{k0}\}$, and by
Eq.~(\ref{eq:redundancies}), we set the values of~$\{[{\bf G}]_{0k}\}$ as
well.  Imposing the Zauner condition fixes $\{[{\bf F}]_{0k}\}$ in terms
of~$\{[{\bf F}]_{k0}\}$.  Specifically, for $k \neq 0$, we have
\begin{equation}
[{\bf F}]_{0k} = [{\bf F}]_{k0} = [{\bf G}]_{k0} = [{\bf G}]_{0k} = \frac{1}{d+1}.
\end{equation}
Recalling that
\begin{equation}
[{\bf F}]_{0k} = \sum_l [{\bf G}]_{lk},
\end{equation}
we therefore find that
\begin{equation}
[{\bf G}]_{0k} = \sum_l [{\bf G}]_{lk}.
\end{equation}
In other words, the Zauner condition implies that if we add up the
entries in a column, leaving out the entry on the top row, they must
all cancel out and leave zero.  We knew already, thanks to the
Wiener--Khinchin theorem and Eq.~(\ref{eq:G-cc}), that the imaginary
parts will sum to zero.  Now we can establish this for the real parts
as well.

The left-most column of ${\bf G}$ is also its top row, which tells us
the \emph{averages} of each column of~${\bf F}$.  So, if the Zauner
orbits leave only one element in a column unspecified, then we can
fill in that element, because we know the average over the whole
column vector.  This proves the $3d$ conjecture in dimension $d =
9$.

We now turn to the simplification that the Zauner conjecture provides
for numerical search efforts.  By postulating that the SIC fiducial we
are looking for is a Zauner eigenvector, we can significantly reduce
the effective size of the search space.  First, suppose that $U$ is a
unitary of order $n$, so that
\begin{equation}
U^n = I,
\end{equation}
and the eigenvalues of $U$ can all be written
\begin{equation}
\lambda_m = \exp\left(\frac{2\pi im}{n}\right),
\end{equation}
with $m$ an integer.  The projector onto the eigenspace with this
eigenvalue is
\begin{equation}
P_m = \frac{1}{n} \sum_{j=0}^{n-1} \frac{1}{\lambda_m^j} U^j.
\end{equation}
We can restrict our numerical search to this subspace by projecting
our vectors into it,
\begin{equation}
\ket{\psi} \to P_m\ket{\psi},
\end{equation}
at each iteration of the optimization algorithm.

Most of the known solutions were found by postulating Zauner symmetry.
Scott has also found several solutions by assuming that the fiducial
was an eigenvector of another Clifford unitary.  For an in-depth
exposition of these variations, see~\cite{Scott:2017}.

\section{Exhaustive Searches}

Suppose that $\ket{\psi_0} \in \hilb_d$ is a vector in a
Weyl--Heisenberg SIC, and let $U$ be a Clifford unitary.  Applying $U$
to the vector $\ket{\psi_0}$ will yield some vector,
\begin{equation}
\ket{\chi_0} = U \ket{\psi_0}.
\end{equation}
The Weyl--Heisenberg orbit of $\ket{\psi_0}$ is a SIC, so what about
the orbit of~$\ket{\chi_0}$ under the same group?  We define
\begin{equation}
\ket{\chi_{kl}} = D_{kl} \ket{\chi_0},
\end{equation}
and we consider the squared magnitudes of the inner products
\begin{equation}
|\braket{\chi_{0}}{\chi_{kl}}|^2
 = |\bra{\psi_0}U^\dag D_{kl} U \ket{\psi_0}|^2.
\end{equation}
Because $U$ is a Clifford unitary, it maps $D_{kl}$ to some
Weyl--Heisenberg operator $D_{k'l'}$, with any phase factor dropping
out when we take the magnitude of the inner product.  So,
\begin{equation}
|\braket{\chi_{0}}{\chi_{kl}}|^2 =
|\braket{\psi_{0}}{\psi_{k'l'}}|^2,
\end{equation}
meaning that the image of our original SIC under the mapping $U$ is
also a SIC.  One way in which the Hesse SIC is remarkable is that it
is invariant under the entire Clifford group.  For contrast, we can
take the vector
\begin{equation}
\ket{\psi_{0}^{(\rm Norrell)}} = \frac{1}{\sqrt{2}}
\left(\begin{array}{c} 0 \\ 1 \\ 1
 \end{array}\right),
\end{equation}
which differs from the Hesse SIC fiducial in
Eq.~(\ref{eq:hesse-fiducial}) by a sign.  The orbit of this vector
under the Clifford group is a set of \emph{four} separate SICs,
comprising 36 vectors in all---the so-called \emph{Norrell states,}
which are significant in the theory of quantum
computation~\cite{Veitch:2014, Stacey:2016c}.

We consider two SICs equivalent if they can be mapped into each other
by a Clifford unitary.  In fact, it is convenient to \emph{extend} the
Clifford group by including the anti-unitary operation of complex
conjugation.  The \emph{extended Clifford group} for dimension $d$,
$\ECgp(d)$, is the set of all unitary and anti-unitary operators that
send the Weyl--Heisenberg group to itself.  For (extensive) details,
we again refer to Appleby~\cite{Appleby:2005b, Appleby:2009}.

In order to search the space as exhaustively as possible and create a
catalogue of all essentially unique SICs, Scott's code chooses initial
vectors at random under the unitarily invariant Haar measure on the
complex projective space $\mathbb{C}P^{d-1}$.  Once enough solutions
are found---generally, this means hundreds of them---the code then
refines their precision, as described above.  Then, we must identify
unique orbits under the extended Clifford group.  This last step is
computationally demanding, because we must translate each solution
vector $\ket{\psi}$ by each element in the extended Clifford group
$\ECgp(d)$.  However, in the process, Scott's algorithm also finds the
stabilizer group of each fiducial, which is important information.
The task of determining when two SICs are equivalent up to a unitary
or anti-unitary transformation has been discussed in depth by
Zhu~\cite{Zhu:2012}, and we expect that additional theoretical
insights may lead to an improved algorithm for this step.

Following this procedure, Scott has carried out exhaustive searches in
dimensions up to $d = 90$.  We strongly expect his catalogue of solutions 
to be complete up to that point:  All Weyl--Heisenberg SICs in those 
dimensions are equivalent to the ones tabulated, up to equivalence under the 
extended Clifford group.

\section{Discussion}

In the preceding sections, we have described the process of finding
SIC fiducial vectors numerically.  However, some patterns among SICs
have only become apparent when \emph{exact} solutions were studied
carefully.  Suppose we refrain from taking the magnitude-squared in
our definition of a SIC, Eq.~(\ref{eq:SIC}).  Then
\begin{equation}
\braket{\psi_j}{\psi_k} = \frac{e^{i\theta_{jk}}}{\sqrt{d+1}},
\end{equation}
for some set of phases $\{e^{i\theta_{jk}}\}$.  (In fact, one can
reconstruct the SIC from knowing the phases~\cite{Appleby:2011a}.)  It
was recently discovered that when $d > 3$, for all the known
Weyl--Heisenberg SICs, these phases have a remarkable meaning in
algebraic number theory: They are \emph{units in ray class fields and
  extensions thereof}~\cite{Appleby:2016}.  This is a topic to which
we can hardly do justice here, and indeed, treatments accessible to
anyone who is not already an algebraic number theorist have only
recently been attempted~\cite{Bengtsson:2016, FOOP:2017}.  For now, we
content ourselves with the observation that this area of number theory
is the territory of Hilbert's twelfth problem, one of the still
outstanding questions on history's most influential list of
mathematical challenges~\cite{Gray:2001}.  (Specialists may recall
that according to the Kronecker--Weber theorem, any abelian extension
of the rationals is contained in a cyclotomic field.  When we instead
consider abelian extensions of real quadratic fields, the analogue of
the cyclotomic fields are the ray class fields.  The phases of
Weyl--Heisenberg SICs appear to be playing a role regarding ray class
fields much like the role that roots of unity play with cyclotomic
fields.  Moreover, recalling Eq.~(\ref{eq:real-inprod}), it is
intriguing that in the real-vector-space version of equiangular lines,
we discard a phase factor that is a unit among the ordinary integers,
while in the complex Weyl--Heisenberg case, the phases turn out to be
units among algebraic integers.) From Hilbert space to Hilbert's
twelfth problem!  What physicist would ever have anticipated that?
And who could turn down the opportunity to intermingle two subjects
that had seemed so widely separated?

SICs have found relevance, not just in quantum computation, but in
signal-processing tasks like high-precision radar~\cite{Howard:2006}
and speech recognition~\cite{Balan:2009}.  In February 2016, our
colleague Marcus Appleby attended a conference in Bonn, Germany on
uses of the Weyl--Heisenberg group.  Many participants were engineers,
including representatives from the automotive and cell-phone
industries.  Appleby was told that if he managed to construct a SIC in
dimension 2048, he should patent it~\cite{Marcus-email}.  At the
moment, dimension 2048 is beyond our abilities for algebraic or
numerical solutions, but this may not always be the case.

\section{Acknowledgments}

We are deeply indebted to Andrew J.\ Scott, the guru of SIC numerical
solutions, for code and for discussions.  We also thank Marcus Appleby
for many conversations, and Gary McConnell for email feedback on the
original arXiv version of this article.  This research was supported
in part by MCH's Oracle Undergraduate Research Fellowship at UMass
Boston.

\section{Author Contributions}

MCH ran the calculations on Chimera to find SICs in dimensions 122
through 151.  BCS wrote the paper.  CAF directed the research,
contributed to the bibliography and worked with BCS in revising
the paper.

\nocite{Ahmadi:2017}
\nocite{Albouy:2007}
\nocite{Andersson:2014}
\nocite{Andersson:2014b}
\nocite{Andersson:2017}
\nocite{Appleby:2005b}
\nocite{Appleby:2007a}
\nocite{Appleby:2009}
\nocite{Appleby:2009a}
\nocite{Appleby:2011}
\nocite{Appleby:2012}
\nocite{Appleby:2012b}
\nocite{Appleby:2013}
\nocite{Aravind:2017}
\nocite{Baldwin:2016}
\nocite{Ballester:2005}
\nocite{BarOn:2009}
\nocite{BarOn:2009b}
\nocite{Belovs:2008}
\nocite{Beneduci:2013}
\nocite{Bengtsson:2009}
\nocite{Bengtsson:2010}
\nocite{Bengtsson:2012}
\nocite{Bengtsson:2012b}
\nocite{Bengtsson:2017}
\nocite{Blanchfield:2014}
\nocite{Bodmann:2007}
\nocite{Bodmann:2008}
\nocite{Bodmann:2010}
\nocite{Brandsen:2016}
\nocite{Bruzda:2017}
\nocite{Cao:2017}
\nocite{Carmeli:2011}
\nocite{Carmeli:2012}
\nocite{Casazza:2017}
\nocite{Sogamoso:2017}
\nocite{Chen:2015}
\nocite{Chien:2013}
\nocite{Chien:2015}
\nocite{Chien:2017}
\nocite{Colin:2005}
\nocite{DallArno:2011}
\nocite{DallArno:2014}
\nocite{DallArno:2014a}
\nocite{DallArno:2015}
\nocite{Dang:2013}
\nocite{Dang:2015}
\nocite{Delsarte:1975}
\nocite{Englert:2004}
\nocite{Et-Taoui:2000}
\nocite{Et-Taoui:2002}
\nocite{Et-Taoui:2016}
\nocite{Ferrie:2008}
\nocite{Ferrie:2009}
\nocite{Feynman:1987}
\nocite{Fickus:2009}
\nocite{Fickus:2012}
\nocite{Filippov:2011}
\nocite{Flammia:2006}
\nocite{Fuchs:2003}
\nocite{Fuchs:2004}
\nocite{Fuchs:2011}
\nocite{Fuchs:2011b}
\nocite{Fuchs:2016b}
\nocite{Fuchs:2017b}
\nocite{Godsil:2009}
\nocite{Gour:2014}
\nocite{Goyeneche:2011}
\nocite{Goyeneche:2014}
\nocite{Goyeneche:2016}
\nocite{Goyeneche:2017}
\nocite{Grassl:2004}
\nocite{Grassl:2005}
\nocite{Grassl:2008}
\nocite{Graydon:2013}
\nocite{Graydon:2016}
\nocite{Graydon:2016a}
\nocite{Graydon:2017}
\nocite{Herman:2009}
\nocite{Howard:2016}
\nocite{Hughston:2007}
\nocite{Hulek:1983}
\nocite{Jedwab:2014}
\nocite{jedwab:2015}
\nocite{Kalev:2012}
\nocite{Kalev:2012a}
\nocite{Kayser:2015}
\nocite{Khatirinejad:2008}
\nocite{MKF:2008}
\nocite{Khrennikov:2017}
\nocite{Kim:2007}
\nocite{Klappenecker:2005}
\nocite{Koenig:1992}
\nocite{Koenig:1999}
\nocite{Koenig:2005}
\nocite{Larsson:2010}
\nocite{Malikiosis:2016}
\nocite{Matthews:2009}
\nocite{Mixon:2012}
\nocite{McConnell:2014}
\nocite{Oreshkov:2011}
\nocite{Oszmaniec:2016}
\nocite{Petz:2012}
\nocite{Petz:2012a}
\nocite{Petz:2014}
\nocite{Planat:2017}
\nocite{Rastegin:2013}
\nocite{Rastegin:2014a}
\nocite{Rehacek:2004}
\nocite{Renes:2004a}
\nocite{Renes:2005}
\nocite{Rosado:2011}
\nocite{Rosado:2013}
\nocite{Saraceno:2016}
\nocite{Scott:2006}
\nocite{Shen:2015}
\nocite{Slomczynski:2014}
\nocite{Slomczynski:2016}
\nocite{Szymusiak:2014}
\nocite{Szymusiak:2015}
\nocite{Szymusiak:2017}
\nocite{Tabia:2012}
\nocite{Tabia:2013}
\nocite{Tabia:2013b}
\nocite{Waldron:2011}
\nocite{Waldron:2013b}
\nocite{Wetering:2017}
\nocite{Wootters:2006}
\nocite{Wootters:2009}
\nocite{Xu:2015}
\nocite{YadsanAppleby:2012}
\nocite{Zhu:2010}
\nocite{Zhu:2010a}
\nocite{Zhu:2011}
\nocite{Zhu:2012}
\nocite{Zhu:2014}
\nocite{Zhu:2015a}
\nocite{Zhu:2015b}
\nocite{Zhu:2016a}

\bibliographystyle{utphys}

\bibliography{sic}

\end{document}